\documentstyle[preprint,aps]{revtex}

\draft

\begin{document}

\title{From quantum theory to classical dynamics under spontaneous wave-packet reduction}

\author{C. F. Huang}

\address{$^{1}$Department of Physics, National Taiwan University, Taipei, Taiwan, R. O. C. \newline
$^{2}$National Measurement Laboratory, Center for Measurement Standards, Industrial Technology Research Institute, Hsinchu, Taiwan 300, R. O. C.}

\date{\today}

\maketitle

\begin{abstract}

A quantum master equation of the Lindblad form is obtained in this paper by considering the spontaneous wave-packet reduction. Different classical equations can be derived after exactly mapping such a quantum master equation to a continuous time random walk (CTRW). Although this CTRW is a quasiclassical walk, the effects due to the quantum interference can still be important in such a walk. Macroscopically, we shall consider the uncertainty of the potential and determine the effective transition probability by a family of Schr\"{o}dinger equations (or operators).

\end{abstract}

\newpage



\section{Introduction}

While classically a particle has the definite position and momentum at any time, in quantum mechanics it is described by a complex wavefunction and its position and momentum cannot be determined simultaneously from the uncertainty principle. \cite{Sakurai,Newton} A wavefunction, in fact, can be a superposition of microscopic wave packets and become very extended with respect to either the momentum or position. To determine the momentum (position), in quantum mechanics it is assumed that the wavefunction collapses in momentum (position) space under the momentum (position) measurements. \cite{Sakurai,Vladimir} With suitable assumptions, the quantum master equation of the Lindblad form \cite{Alfredo,Lindblad} can be used to explain the appearance of the classical world. \cite{Giulini} In addition to the quantum Liouville term, in such an equation there is a relaxation term which can induce the collapse of states or reduction of wave packets \cite{Sakurai,Vladimir,Giulini,Michael,Wiseman}. The relaxation term can be due to the quantum measurements or the interactions with the reservoir. \cite{Giulini,Michael,Scott,Huang,Kampen} Although no state has the definite momentum and position simultaneously, a wavefunction can collapse to a wave packet to have the momentum and position within a small uncertainty under the phase-space measurement \cite{Scott,Braunstein}. A classical particle, in fact, is described by a microscopic wave packet of which the uncertainties in both the momentum and space are negligible macroscopically. \cite{Sakurai} Let $|\phi \rangle $ be a normalized microscopic wave packet located at the position ${\bf x}=0$ with the momentum ${\bf p}=0$. Then in the set
\begin{equation}
S_{\phi }\equiv \{|\phi _{ {\bf x}_{0}, {\bf p}_{0}}\rangle =T_{ {\bf x}
_{0}} {\cal T} _{ {\bf p}_{0} }|\phi \rangle \},
\end{equation}
each ket is a wave packet located at $ {\bf x}= {\bf x}_{0}$ with ${\bf p}={\bf p}_{0}$. Here $T_{ {\bf x}_{0} }$ is the translation operator to shift each ket by the displacement ${\bf x}_{0}$, and ${\cal T}_{ {\bf p}_{0}}$ is the operator to shift each ket by the momentum $ {\bf p}_{0}$. 

In this paper, a particle moving in $R^{n}$ space under the Hamiltonian
\begin{equation}
H=\frac{ {\bf p}^{2} }{2m}+V( {\bf x} )
\end{equation}
is considered, where $n$ is an integer, $m$ is the mass of the particle, and $V({\bf x})$ is the potential. Let $\rho (t)$ be the density matrix at time $t$ and satisfies the normalized condition
\begin{equation}
tr\rho (t)=1.
\end{equation}
(All the quantities and equations are discussed in the Schr\"{o}dinger picture \cite{Sakurai} in this paper.) Considering a lifetime $\tau $ for the particle to become wave packets in $S_{\phi }$ spontaneously, in section II the equation
\begin{equation}
\frac{\partial }{\partial t}\rho (t)=\frac{i}{\hbar }[\rho (t),H]-\frac{\rho
(t)}{\tau }+\frac{1}{\tau }\int \frac{d^{n}xd^{n}p}{(2\pi \hbar )^{n}}|\phi
_{{\bf x}, {\bf p}}\rangle \langle \phi _{ {\bf x}, {\bf p}}|\rho
(t)|\phi _{ {\bf x}, {\bf p} }\rangle \langle \phi _{ {\bf x}, {\bf p}
}|.
\end{equation}
is derived under suitable assumptions, where $\hbar $ is the Plank constant divided by $2\pi $. Eq. (4), in fact, is a quantum master equation of the Lindblad form and can be exactly mapped to a continuous time random walk (CTRW) \cite{Haus,Montroll} in which the probability density
\begin{equation}
r({\bf x}, {\bf p},t) \equiv \frac{1}{\tau }\langle \phi _{ {\bf x},
{\bf p} } |\rho (t)| \phi _{ {\bf x}, {\bf p} } \rangle .
\end{equation}
In such a CTRW, $r({\bf x}, {\bf p},t)$ is governed by
\begin{equation}
r({\bf x}, {\bf p},t)=\int_{t_{0}}^{t}dt^{\prime }\int \frac{
d^{n}x^{\prime }d^{n}p^{\prime }}{(2\pi \hbar )^{n}}\psi ({\bf x}, {\bf p}
; {\bf x}^{\prime },{\bf p}^{\prime };t-t^{\prime })r( {\bf x}
^{\prime }, {\bf p}^{\prime },t^{\prime })+s({\bf x}, {\bf p},t).
\end{equation}
with the transition probability density (memory kernel) 
\begin{equation}
\psi ({\bf x}, {\bf p}; {\bf x}^{\prime }, {\bf p} ^{\prime}
;t-t^{\prime })\equiv \frac{1}{\tau }e^{-(t-t^{\prime })/\tau }|\langle
\phi _{ {\bf x}, {\bf p}}|e^{-\frac{i}{\hbar }(t-t^{\prime })H}|\phi _{
{\bf x}^{\prime },{\bf p}^{\prime }}\rangle |^{2}
\end{equation}
and the source term
\begin{equation}
s({\bf x}, {\bf p},t)\equiv \frac{1}{\tau }e^{-(t-t _{0}) / \tau
}\langle \phi _{ {\bf x}, {\bf p} }| e ^{-\frac{i}{\hbar }(t-t _{0})H}
\rho (t_{0}) e^{ \frac{i}{\hbar }(t-t _{0})H} 
|\phi _{ {\bf x}, {\bf p}}\rangle .
\end{equation}
Here $t_{0}$ is the initial time, and in Eq. (6) $\int \frac{d^{n}x^{\prime }d^{n}p^{\prime }}{(2\pi \hbar )^{n}}$ presents the sum over all possible states in the CTRW with $(2 \pi \hbar) ^{-n}$ as the density of states in phase space. Since $r( {\bf x}, {\bf p},t)$ is a real function defined in the (classical) phase space at any time $t$, such a CTRW can be taken as a quasiclassical walk. As shown in section III, however, the effects due to the quantum interference can be important even in the macroscopic scale. 

In the quantum master equation of the Lindblad form, the irreversibility is due to the relaxation term. On the other hand, E. T. Jaynes \cite{Jaynes} obtained a theory for irreversible processes by considering a family of Schr\"{o}dinger equations (or operators) under the assumption that the potential may not be known certainly. The random Schr\"{o}dinger equations (or operators) \cite{Laszlo,Norbert,Ivan}, in which the potentials are random fields, have been used to study
phenomena due to the quantum interference such as the Anderson localization \cite{Ivan,Anderson}. Macroscopically, as shown in section III, we shall consider the uncertainty on the position of the potential and determine the effective transition rate by a family of the Schr\"{o}dinger equations (or operators). With suitable assumptions, the CTRW can correspond to the classical Liouville equation \cite{Liboff} or classical master equation \cite{Huang,Liboff}. 

Different quantum approcahes have been developed to derive the classical equations. \cite{Giulini,Jaynes,Laszlo,Liboff,Swenson,Gough,Michael2,Smirnov} It is shown in section IV that Eq. (4) can be related to the momentum-position collapse model discussed in Refs. \cite{Michael2} and \cite{Benatti}. The density $\psi ({\bf x}, {\bf p}; {\bf x}^{\prime }, {\bf p} ^{\prime} ;t-t^{\prime })$ given by Eq. (7) is the product of a temporal factor and the well-known quantum transition probability for being $|\phi _{{\bf x}^{\prime},{\bf p}^{\prime}} \rangle $ at time $t^{\prime}$ to become $| \phi _{ {\bf x}, {\bf p} } \rangle$ at time $t$. Therefore, the CTRW provides an intuitive way to relate the quantum theory to the classical motions under the wave-packet reduction. The transition probability density  $\psi ({\bf x}, {\bf p}; {\bf x}^{\prime }, {\bf p} ^{\prime} ;t-t^{\prime })$, in fact, is proportional to the jump rate in the corresponding quantum trajectory formulation. \cite{Minami,Breuer} The conclusion is made in section VI.

\section{Quantum measurements, reduction of wave packets, and continuous time random walk}

Under the (quantum) Liouville flow with the Hamiltonian $H$, the time-dependent density matrix satisfies \cite{Sakurai,Kampen}
\begin{equation}
\rho (t+\Delta t)=e^{-\frac{i}{\hbar}\Delta tH}\rho (t)e^{\frac{i}{\hbar} \Delta tH}.
\end{equation}
But if we performed a measurement of the nondegenerate observable $A$ at $t+\Delta t$, the particle must collapse suddenly to an eigenket of $A$ at $t+\Delta t$. \cite{Sakurai} Let
\begin{equation}
{\LARGE \{ } |a \rangle | \text{ } A|a\rangle = a |a \rangle , \langle a | a \rangle = 1 {\LARGE \} }.
\end{equation}
be the set of normalized eigenkets of $A$. (Note that the eigenkets of momentum or position operators cannot be normalized in $L _{2}$. \cite{Sakurai}) Since the probability to collapse to $|a\rangle $ is
\begin{equation}
c_{a} (t,\Delta t) \equiv \langle a|e^{-\frac{i}{\hbar }\Delta tH} \rho (t) e^{\frac{i}{\hbar }\Delta tH}|a\rangle ,
\end{equation}
at $t+\Delta t$ we shall replace Eq. (9) by 
\begin{equation}
\rho (t+\Delta t)=\sum_{a}c_{a} (t, \Delta t) |a\rangle \langle a|
\end{equation}
under such a measurement. \cite{Sakurai}

Now assume that the particle has the lifetime $\tau $ to become to an eigenfunction of $A$ spontaneously rather than collapsing suddenly. Then during small $\Delta t$ the probabilities to collapse and not to collapse are $\Delta t/\tau $ and $1-\Delta t/\tau $, respectively, and at $t+\Delta t$ we shall set
\begin{eqnarray}
\rho (t+\Delta t) &=&(1-\Delta t/\tau )e^{-\frac{i}{\hbar }\Delta tH}\rho
(t)e^{\frac{i}{\hbar }\Delta tH}
\end{eqnarray}
\[
+(\Delta t/\tau )\sum_{a} c_{a} (t,\Delta t) |a\rangle \langle a|+o(\Delta t^{2}).
\]
At the right hand side of the above equation, the first term presents the factor due to the Liouville flow and the second term presents the factor due to the collapse. As $\Delta t \rightarrow 0$, with Eq. (11) we can reduce Eq. (13) as
\begin{equation}
\frac{\partial }{\partial t}\rho (t)=\frac{i}{\hbar }[\rho (t),H]-\frac{\rho
(t)}{\tau }+\frac{1}{\tau }\sum_{a}|a\rangle \langle a|\rho (t)|a\rangle
\langle a|.
\end{equation}
It is easy to check that $tr\rho (t)$ is kept under the above equation. With some calculations, we can see that the last
two terms in Eq. (14) induce the decay of phases (off-diagonal terms) with respect to the eigenkets of $A$ without affecting the diagonal terms.

Although a wavefunction cannot collapse to a state with the definite momentum and position from the uncertainty principle, it can collapse to a ket in $S_{\phi }$ defined in Eq. (1). To replace eigenstates of $A$ by the states in $S_{\phi }$, the third term of Eq. (14) should be replaced by $c\int \frac{d^{n}xd^{n}p}{(2\pi \hbar )^{n}}|\phi _{{\bf x},{\bf p} } \rangle \langle \phi _{ {\bf x}, {\bf p} } | \rho (t)|\phi _{ {\bf x}, {\bf p} } \rangle \langle \phi _{ {\bf x}, {\bf p} }|$ with $c$ as a constant. (We shall replace $\sum_{a}$ by $\int \frac{d^{n}xd^{n}p}{(2\pi \hbar )^{n}}$ because the momentums and positions are continuous parameters.) To preserve $tr\rho (t)$, we shall set $c=1/\tau $
and Eq. (4) is obtained. When the spatial width of $|\phi \rangle $ tends to zero, each ket in $S_{\phi }$ becomes an eigenket of the position operators and hence Eq. (4) describes the collapse in position. On the other hand, Eq. (4) describes the collapse in the momentum space as the width of $|\phi \rangle $ in momentum shrinks to zero.

The density matrix $\rho (t)$ governed by Eq. (4), in fact, satisfies 
\begin{eqnarray}
\rho (t) = e^{-(t-t_{0})/\tau }e^{-\frac{i}{\hbar }(t-t_{0})H}\rho (t_{0})e^{\frac{i}{\hbar }(t-t_{0})H}+
\end{eqnarray}
\[
\int_{t_{0}}^{t}dt^{\prime} e^{-(t-t^{\prime })/ \tau }  \int \frac{ d^{n}xd^{n}p}{(2\pi \hbar )^{n}}r({\bf x},{\bf p},t 
^{\prime})e^{-\frac{i}{\hbar}(t-t^{\prime })H}|\phi _{ {\bf x}, {\bf p} } \rangle \langle \phi _{{\bf x}, {\bf p} } | e 
^{\frac{i}{\hbar }(t-t^{\prime })H}
\]
with $r({\bf x},{\bf p},t)$ defined in Eq. (5). To obtain the above equation intuitively, note that the first term at the right hand side of Eq. (15) corresponds to the case that the particle moves without collapsing. The matrix $e^{-\frac{i}{\hbar }(t-t_{0})H}\rho (t_{0})e^{\frac{i}{\hbar} (t-t_{0})H}$ corresponds to the quantum Liouville flow and the factor
$e^{-(t-t_{0})/\tau }$ is the probability for the particle not to collapse. On the other hand, if the particle collapses and the corresponding density matrix becomes $|\phi _{ {\bf x},{\bf p} } \rangle \langle \phi _{ {\bf x} , {\bf p} } | $ at time $t^{\prime }$, at time $t > t ^{\prime }$ the density matrix is $e^{-\frac{i}{\hbar }(t-t^{\prime })H}|\phi _{ {\bf x}, {\bf p} } \rangle \langle \phi _{ {\bf x}, {\bf p} } | e ^{ \frac{i}{\hbar } (t-t^{\prime }) H } $ under the quantum Liouville flow. The probability to become $|\phi _{ {\bf x}, {\bf p} } \rangle \langle \phi _{ {\bf x}, {\bf p} } | $ at $t^{\prime }$ is proportional to $r( {\bf x}, {\bf p}, t ^{\prime})=\frac{1}{\tau }\langle \phi 
_{ {\bf x},{\bf p} } |\rho (t)| \phi _{ {\bf x}, {\bf p} } \rangle$ and the collapse can still occur when $t>t^{\prime }$, so we can expect the last term in Eq. (15). Because Eq. (4) is a first-order differential equation with respect to $t$, to prove Eq. (15) we just need to take the time derivative on the above equation and check the initial condition. Then inserting the above equation into the right hand side of Eq. (5), we can obtain Eq. (6) with the transition probability density and source term defined in Eqs. (7) and (8). 

As shown in Appendix A, actually Eq. (4) can be taken as a particular case of the following equation
\begin{eqnarray}
\frac{\partial }{\partial t}\rho (t) = \frac{i}{\hbar }[\rho (t),{\cal H}
(t)]-\frac{1}{2}\sum_{ll^{\prime }}w_{ll^{\prime }}\{\rho (t),|l\rangle
\langle l|\} +\sum_{ll^{\prime }}w_{ll^{\prime }}|l^{\prime }\rangle \langle 
l|\rho (t)|l\rangle \langle l^{\prime }|.
\end{eqnarray}
Here ${\cal H} (t)$ is a time-dependent Hamiltonian, all states in the last two terms are from a set $S$ composed of normalized kets, the coefficients $w_{ll^{\prime }}$ are nonnegative real numbers, and $\{ A, B \} \equiv AB+BA$ for any two operators $A$ and $B$. The above equation is just the master equation of the Lindblad form
\begin{eqnarray}
\frac{\partial}{\partial t} \rho (t) = \frac{i}{\hbar} [ \rho (t), {\cal H} (t) ] - \frac{1}{2} \sum_{\alpha} \{ \rho (t),  {\cal V} _{\alpha}^{\dagger}  {\cal V} _{\alpha} \} + \sum _{\alpha} {\cal V} 
_{\alpha} \rho (t) {\cal V} _{\alpha} ^{\dagger}
\end{eqnarray}
with $\alpha \rightarrow (l,l^{\prime})$ and the operator ${\cal V}_{\alpha} \rightarrow w _{ll^{\prime}} ^{1/2} | l^{\prime} \rangle \langle l |$. It is shown in Ref. \cite{Huang} that Eq. (16) can be mapped to an extended random walk, and Eq. (15) in this paper can be obtained from Eq. (13) in Ref. \cite{Huang}, in fact. (But note that in Ref. \cite{Huang} the Hamiltonian is time-dependent and the transition probability density is not necessary of the form $\psi (l^{\prime },l,t-t^{\prime })$.) To prove that Eq. (6) defines a CTRW, we need to show that \cite{Huang,Haus,Montroll} 
\begin{equation}
\int_{t^{\prime }}^{\infty }dt\int \frac{d^{n}xd^{n}p}{(2\pi \hbar )^{n}} \psi ({\bf x}, {\bf p}; {\bf x}^{\prime }, {\bf p}^{\prime}; t-t^{\prime })=1.
\end{equation}
\begin{equation}
\int_{t_{0}}^{\infty }dt\int \frac{d^{n}xd^{n}p}{(2\pi \hbar )^{n}}s( {\bf x}, {\bf p},t)=1.
\end{equation}
The proof is given in Appendix A.

\section{From quasiclassical CTRW to classical equations of motion}

By considering a lifetime $\tau$ for the particle to become a microscopic wave packet spontaneously, in the last section Eq. (4) is derived and is mapped to a CTRW no matter how long the lifetime $\tau $ is. In this section, it will be shown that macroscopically such a CTRW can be reduced to the classical Liouville equation, classical master equation, or a random walk exhibiting effects due to the quantum interference. (It should be emphasized that some assumptions used in this section do not hold in the conventional quantum scattering theory. \cite{Sakurai,Newton})

Because the source term given by Eq. (8) contains a decay factor $e^{-(t-t_{0} )/ \tau}$, such a source term is unimportant when $t>>t_{0}$ and the statistical properties of the CTRW are determined by the transition probability density in Eq. (7). Define
\begin{equation}
G({\bf x},{\bf p};{\bf x}^{\prime},{\bf p}^{\prime }; \xi ) \equiv | \langle \phi _{ {\bf x}, {\bf p} } | e ^{ -\frac{i}{\hbar} \xi H} | \phi _{{\bf x}^{\prime }, {\bf p}^{\prime } } \rangle | ^{2}
\end{equation}
as the expectation value of the time-dependent ket $|\phi _{ {\bf x} ^{\prime }, {\bf p}^{\prime} } (t) \rangle \equiv e^{ -\frac{i}{\hbar} tH } | \phi _{ {\bf x}^{\prime}, {\bf p}^{\prime}} \rangle $ in ket $ | \phi _{ {\bf x}, {\bf p} } \rangle $ at $t=\xi $. We can rewrite the transition probability density as
\begin{equation}
\psi ( {\bf x}, {\bf p}; {\bf x}^{\prime }, {\bf p}^{\prime };\xi )= \frac{1}{\tau }e^{-\xi /\tau }G({\bf x}, {\bf p}; {\bf x}^{\prime },{\bf p}^{\prime}; \xi )
\end{equation}
with $\xi =t-t^{\prime }$. The time-dependent ket $|\phi _{ {\bf x} ^{\prime}, {\bf p}^{\prime} } (t) \rangle $, in fact, is governed by the Schr\"{o}dinger equation
\begin{equation}
i\hbar \frac{\partial }{\partial t}|\phi _{ {\bf x}^{\prime }, {\bf p} ^{\prime }}(t)\rangle =H|\phi 
_{ {\bf x}^{\prime }, {\bf p}^{\prime} } (t)\rangle .
\end{equation}
To determine the behavior of the CTRW, we only need to discuss $G({\bf x}, {\bf p}; {\bf x}^{\prime }, {\bf p} ^{\prime };\xi )$ or $|\phi _{ {\bf x}^{\prime }, {\bf p} ^{\prime} } ( \xi ) \rangle $ when $ \xi $ is comparable with $\tau $ because of the factor $\frac{1}{\tau } e^{- \xi /\tau }$ at the right hand side of Eq. (21).

Although the CTRW defined by Eqs. (5)-(8) is a walk over the phase space and can be taken as a quasiclassical random walk, the effects due to the quantum interference may still be important in such a CTRW. To see this, let $V({\bf x} )$ in Eq. (2) be a random potential. While classically a particle with nonzero velocity can move to infinity if $V( {\bf x})$ is very weak, the quantum interference can make $|\phi _{ {\bf x}^{\prime }, {\bf p} ^{\prime} } (t) \rangle $ governed by Eq. (22) become immobile after passing long enough time. \cite{Anderson} In the one or two-dimensional cases, in fact, insulating behaviors are expected no matter how weak the random potential is. \cite{Queiroz,Huang2}  Since $G( {\bf x}, {\bf p}; {\bf x} ^{\prime}, {\bf p}^{\prime };\xi )$ is the expectation value of $|\phi _{ {\bf x}^{\prime },{\bf p}^{\prime} } ( \xi ) \rangle $ in $|\phi _{ {\bf x} , {\bf p} } \rangle $ and the behavior of the CTRW is determined by $G({\bf x}, {\bf p};{\bf x}^{\prime},{\bf p}^{\prime};\xi )$ when $\xi $ is comparable with $\tau $, we can expect that the effects due to the quantum interference exist if $\tau $ is very long. It should be noted that in Eq. (15) $r({\bf x},{\bf p},t^{\prime})$ is followed by the matrix $e^{-\frac{i}{\hbar }(t-t^{\prime })H}|\phi _{ {\bf x}, {\bf p}}\rangle \langle \phi _{ {\bf x}, {\bf p} } | e ^{ \frac{i}{\hbar}(t-t^{\prime})H}$, which is an immobile state rather than a state with the velocity ${\bf p} /m $ when $t-t^{\prime }$ is large. Such an immoblie state is the superposition of the microscopic states in $S _{\phi}$, and can be a wave packet of macroscopic scale. On the other hand, consider the case that in Eq. (2) the change of $V({\bf x})$ is insignificant in the quantum scale and $m$ is very large. In such a case, it is known that Eq. (22) can correspond to the classical Liouville equation for a very long time. Hence we can expect that with suitable $\tau $, the CTRW corresponds to the Liouville equation in such a scale. The
details are given in Appendix B.

It is shown in Appendix C that under suitable assumptions, macroscopically the effective transition probability density of the CTRW is
\begin{eqnarray}
\Psi ({\bf x}, {\bf p}; {\bf x}^{\prime }, {\bf p}^{\prime}; t-t^{\prime }) = \frac{1}{\Omega _{1} \Omega _{2}} \int_{ | {\bf y} | <\Delta _{1} } d ^{n} y \int _{ | {\bf k} | < \Delta _{2} } d ^{n} k \psi ( {\bf x} + {\bf y}, {\bf p}+{\bf k};{\bf x}^{\prime}+ {\bf y},{\bf p} ^{\prime }+ {\bf k}; t-t
^{\prime})  
\end{eqnarray}
Here $\Delta _{1}$ and $\Delta _{2}$ are real parameters so that two points $({\bf x}_{1},{\bf p}_{1})$ and $({\bf x}_{2},{\bf p}_{2})$ in the phase space are too close to be distinguished macroscopically if $|{\bf x}_{1}-{\bf x}_{2}|< \Delta _{1}$ and $| {\bf p}_{1}-{\bf p} _{2}|<\Delta _{2}$, and $\Omega _{1}$ and $\Omega _{2}$ present the volumes $\int _{|{\bf y}| < \Delta _{1}} d ^{n} y$ and $\int _{|{\bf p}| < \Delta _{2}} d ^{n} p$. That is, the macroscopic jump rate is the average of the microscopic transition probability on a small volume which is taken as a point macroscopically. Define $H _{ {\bf y} } \equiv \frac{ {\bf p}^{2} }{2m}+ V({\bf x} - {\bf y} )=T_{ {\bf y} }^{\dagger}HT_{
{\bf y}}$ with ${\bf y} \in R ^{n} $. The density $\Psi ( {\bf x},{\bf p};{\bf x} ^{\prime},{\bf p}^{\prime };t-t^{\prime })$, in fact, is determined by the set $\{ H _{ {\bf y} } , |{\bf y}| < \Delta _{1} \}$ and hence can be related to the irreversible theory suggested by E. T. Jaynes in Ref. \cite{Jaynes}. To see this, note that Eq. (23) can be rewritten as
\begin{equation}
\Psi ({\bf x}, {\bf p}; {\bf x}^{\prime },{\bf p}^{\prime};t-t^{\prime }) =\frac{e^{-(t-t^{\prime })/\tau }}{\tau \Omega _{2}}\int_{|{\bf k}|<\Delta _{2}} \langle \phi _{ {\bf x}, {\bf p} + {\bf k} }|\sigma _{ {\bf x} ^{\prime}, {\bf p} ^{\prime}+{\bf k} }(t - t ^{\prime} )|\phi _{ {\bf x},{\bf p} + {\bf k} } \rangle d^{n} k
\end{equation}
if we define
\begin{equation}
\sigma _{ {\bf x},{\bf p} }( \xi ) \equiv \int d^{n}yF({\bf y})U ^{({\bf y})} (\xi) \sigma ^{(0)} 
_{{\bf x}, {\bf p}} U ^{ ( {\bf y} ) \dagger } (\xi) .
\end{equation}
with $\sigma ^{(0)} _{{\bf x}, {\bf p}} \equiv | \phi _{ {\bf x}, {\bf p} } \rangle \langle \phi 
_{ {\bf x},{\bf p}} |$. Here the real function $F( {\bf y} )$ equals $\Omega _{1} ^{-1}$ for $|{\bf y}| < \Delta _{1}$ and equals 0 for $|{\bf y}| \geq \Delta _{1}$, and $U ^{({\bf y})} (\xi) \equiv e ^{-\frac{i}{\hbar}\xi H_{{\bf y}} }$ with $\xi$ as a real parameter. Eq. (25) can correspond to Eq. (12.5) in Ref. \cite{Jaynes} after taking $F({\bf y}) d^{n} y$ as the probability to move the matrix $\sigma 
^{(0)} _{{\bf x}, {\bf p}}$ by the unitary operator $U ^{({\bf y})} (\xi)$. To prove Eq. (24), note that from the definition of $S _{\phi}$
\begin{equation}
|\phi _{ {\bf x}+{\bf y}, {\bf p} } \rangle =T_{ {\bf x}+{\bf y} } {\cal T} _{ {\bf p} } | \phi \rangle = T _{ {\bf y} }| \phi _{ {\bf x}, {\bf p} } \rangle 
\end{equation}
and from the Taylor's expansion 
\begin{eqnarray}
e^{ -\frac{i}{\hbar}(t-t^{\prime })H_{ {\bf y} } } =\sum_{j} \frac{1}{j!}(-\frac{i}{\hbar}) ^{j} (t-t
^{\prime}) ^{j} (T_{ {\bf y} } ^{\dagger} H T_{ {\bf y} } ) ^{j} 
\end{eqnarray}
\[
=T_{ {\bf y} }^{\dagger} [ \sum_{j} \frac{1}{j!}( - \frac{i}{\hbar } ) ^{j} (t-t^{\prime}) ^{j} H^{j} ] T_{ {\bf y} } = T _{ {\bf y} }^{\dagger}e^{-\frac{i}{\hbar}(t-t^{\prime })H}T_{ {\bf y} }.
\]
Since each $H_{ {\bf y} }$ is obtained from $H$ by performing the position translation on the potential, here the randomness of the time evolution comes from the uncertainty of the potential in the macroscopic scale.

As mentioned above, with suitable assumptions on $\tau$, $m$, and $V({\bf x})$, the transition probability density $\psi$ and the statistical properties of the CTRW can be determined by the classical Liouville equation. If $V( {\bf x} )$ is of the macroscopic scale so that $V( {\bf x}) \simeq V( {\bf x}+ {\bf y} )$ when $| {\bf y} | \leq \Delta _{1}$, it is easy to see that macroscopically the derived CTRW still corresponds to the classical Liouville equation. But if $V({\bf x})$ is random in the scale of $\Delta _{1}$, in the two or three dimensional cases the CTRW may correspond to the classical master equation in a large scale since classically the (microscopic) Liouville equation can be reduced to the classical master equation. \cite{Liboff} On the other hand, in the one or two dimensional cases insulating behaviors due 
to the quantum interference are still expected macroscopically when $V({\bf x})$ is random and $\tau$ is long enough for the quantum interference. Therefore, macroscopically the CTRW defined by Eqs. (5)-(8) may correspond to the Liouville equation, master equation, or a random walk exhibiting the effects due to the quantum interference.

\section{Discussions}

Different quantum approaches have been developed to obtain classical equations. 
\cite{Giulini,Jaynes,Laszlo,Liboff,Swenson,Gough,Michael2,Smirnov} In this paper, Eq. (4) is derived by considering a relaxation time for wavefunctions to collapse to microscopic wave packets, which are concentrated in both momentum and position. In Refs. \cite{Michael2} and \cite{Benatti}, a spontaneous momentum-position localization model is discussed based on the equation of the following form  
\begin{eqnarray}
\frac{\partial}{\partial t} \rho (t) = \frac{i}{\hbar} [ \rho (t), H ] - \frac{ \rho (t) }{\tau}  + c \int \frac{ d ^{n} y d ^{n} k }{(2 \pi \hbar) ^{n}} e ^{- [ \alpha ({\bf p}-{\bf k}) ^{2} + \beta ( {\bf x} -{\bf y})^2 ] } \rho (t) e ^{- [ \alpha ({\bf p}-{\bf k}) ^{2} + \beta ( {\bf x} -{\bf y})^2 ] }.
\end{eqnarray}
Here $\alpha$ and $\beta$ are two positive real parameters, $\tau$ is the relaxation time, and $c$ is a parameter which can be obtained by considering the conservation of $tr \rho (t)$. (Thus $c$ is determined by $\alpha$, $\beta$, and $\tau$.) Let $S_{ {\bf y}, {\bf k} } ^{\alpha , \beta } \equiv \alpha ({\bf p}-{\bf k}) ^{2} + \beta ( {\bf x} -{\bf y})^2 $. We can rewrite Eq. (28) as
\begin{eqnarray}
\frac{\partial}{\partial t} \rho (t) = \frac{i}{\hbar} [ \rho (t), H ] - \frac{ \rho (t) }{\tau}  + c \int \frac{ d ^{n} y d ^{n} k }{(2 \pi \hbar) ^{n}} e ^{- S_{ {\bf y}, {\bf k} } ^{ \alpha , \beta } } \rho (t) e ^{- S_{ {\bf y}, {\bf k} } ^{ \alpha , \beta } }
\end{eqnarray} 
Define $h_{ \bf y} ^{\alpha, \beta} =  \alpha {\bf p} ^{2} + \beta ( {\bf x} -{\bf y})^2  $ as the Hamiltonian of a simple harmonic motion. Then $S_{ {\bf y}, {\bf k} }^{ \alpha , \beta } $ can be obtained by shifting $h_{ \bf y} ^{\alpha, \beta}$ with the momentum translation operator. \cite{Michael2} Thus we can obtain the eigenkets and eigenvalues of $ S_{ {\bf y}, {\bf k} }^{ \alpha , \beta }$ and $exp(-S_{ {\bf y} ,{\bf k} }^{ \alpha , \beta })$ exactly. Let $| E _{ {\bf y}, {\bf k} } ^{\alpha ,\beta} \rangle $ be the ket satisfying $S_{ {\bf y} ,{\bf k} }^{ \alpha , \beta }| E _{ {\bf y}, {\bf k} } ^{\alpha ,\beta} \rangle = E _{ {\bf y}, {\bf k} } ^{\alpha ,\beta} | E _{ {\bf y}, {\bf k} } ^{\alpha ,\beta} \rangle$. Eq. (4), in fact, can be obtained from the above equation by considering the limit 
\begin{eqnarray}
\alpha \rightarrow \infty \text{ and } \beta / \alpha = const.
\end{eqnarray}
In such a limit, each $| E _{ {\bf y}, {\bf k} } ^{\alpha ,\beta} \rangle$ is unchanged while $E _{ {\bf y}, {\bf k} } ^{\alpha ,\beta} \rightarrow \infty$. And it is easy to see that    
\begin{eqnarray}
exp(-S_{ {\bf y},{\bf k} }^{ \alpha , \beta }) = \sum  e^{- E _{{\bf y}, {\bf k} } ^{\alpha , \beta} }  | E _{ {\bf y}, {\bf k} } ^{\alpha ,\beta} \rangle \langle E _{ {\bf y}, {\bf k} } ^{\alpha ,\beta} |
\end{eqnarray}
\[
\rightarrow e^{- {\cal E}_{{\bf y}, {\bf k} } ^{\alpha , \beta} }  | {\cal E}_{ {\bf y}, {\bf k} } ^{\alpha ,\beta} \rangle \langle {\cal E} _{ {\bf y}, {\bf k} } ^{\alpha ,\beta} |,
\] 
where $ {\cal E} _{ {\bf y} , {\bf k} } ^{ \alpha , \beta} $ is the lowest eigenvalue of $S _{ {\bf y} , {\bf k} } ^{ \alpha , \beta }$. Each ket $| {\cal E} _{ {\bf y}, {\bf k} } ^{\alpha , \beta} \rangle$,  in fact, is just a Gaussian wave packet, and Eq. (4) is obtained after setting $| \phi \rangle$ as a wave packet of Gaussian type. 

The quantum trajectory formulation is a powerful numerical approach to solve Eq. (17), the master equation of the Lindblad form. \cite{Minami,Breuer} In such a formulation, stochastic quantum jumps interrupt the continuous evolution determined by the nonhermitian effective Hamiltonian,
\begin{eqnarray}
H _{eff} = H - \frac{i}{2} \sum _{\alpha} {\cal V} _{\alpha} ^{\dagger} {\cal V} _{\alpha}.
\end{eqnarray}
The continuous evolution is responsible for the first two terms at the right hand side of Eq. (17) while the last term in such an equation is interpreted as the origin of quantum jumps. Each ${\cal V} 
_{\alpha}$ is the "collapse" operator to induce the collapse of an arbitrary ket $| a \rangle $ to $ {\cal V} _{\alpha} | a \rangle $ in a quantum jump. For Eq.(4), the effective Hamiltonian is reduced as    
\begin{eqnarray}
H _{eff} = H - i I/2.
\end{eqnarray} 
In addition, we can set $\alpha = ({\bf x},{\bf p}) $ and take each
\begin{eqnarray}
\frac{1}{\tau ^{1/2}} | \phi _{ {\bf x}, {\bf p} } \rangle \langle \phi _{ {\bf x}, {\bf p} } | 
\end{eqnarray}
as a collapse operator. In a trajectory, a wavefunction becomes a wave packet at a quantum jump when each collapse operator is given by the above equation. After the first jump, it is easy to see that the following (conditional) jump rate is proportional to the density $\psi$ given in Eq. (7) with $H _{eff}$ defined in Eq. (33). Therefore, we just need to consider the jump rate proportional to the transition probability density for the CTRW to determine statistical properties.

We can see that $r( {\bf x}, {\bf p}, t) $ for the discussed CTRW is positive definite from its definition while the Wigner distribution \cite{Liboff}, which is taken as the classical correspondence of $\rho (t)$, may contain the negative part. The density $r( {\bf x}, {\bf p} ,t )$, in fact, only records the probability for a wavefunction to become to a wave packet at time $t$. Therefore, its physical meansing is different from that of Wigner distribution. 

\section{Conclusion}

In this paper a quantum master equation of the Lindblad form is obtained by considering the spontaneous wave-packet reduction in phase space. Such an equation can be mapped to a quasiclassical continuous time random walk, from which the classical master equation, classical Liouville equation, or a walk exhibiting effects due to the quantum interference are obtained in the macroscopic scale. The macroscopic transition rate, in fact, is determined by a family of Schr\"{o}dinger operators.

\section*{Acknowledgment}

The author thanks C. C. Chang for his valuable discussions.

\section*{Appendix A}

In Eq. (16), the transitions are over a set of normalized kets and it is not necessary to ask that such a set is composed of orthogonal kets. \cite{Huang} Setting
\begin{equation}
w_{ll^{\prime }}=\delta _{ll^{\prime }}/\tau ,
\end{equation}
Eq. (16) can be reduced to
\begin{eqnarray}
\frac{\partial }{\partial t}\rho (t) = \frac{i}{\hbar }[\rho (t), {\cal H} (t)]-\frac{1}{2 \tau} 
\{ \rho (t) , \sum _{l} | l \rangle \langle l| \} + \frac{1}{\tau }\sum_{l}|l\rangle \langle l|\rho (t)|l\rangle \langle l|.
\end{eqnarray}
In the case that all transitions are over the set $S_{\phi}$, in the above equation each $l$ should correspond to a wave packet $\phi _{{\bf x},{\bf p}}$ and we shall replace $\sum _{l}$ by $\int \frac{ d ^{n} x d ^{n} p}{(2\pi \hbar) ^{n}}$. After some calculations, we can obtain
\begin{eqnarray}
\int \frac{ d ^{n} x d ^{n} p}{( 2 \pi \hbar ) ^{n} } | \phi _{ {\bf x} , {\bf p} } \rangle \langle \phi _{ {\bf x}, {\bf p} } | = I 
\end{eqnarray}
and then derive Eq. (4) from Eq. (36) by setting ${\cal H} (t) =H$, where $I$ is the identity operator. Therefore, Eq. (4) is just a particular case of Eq. (16).

To prove Eqs. (18) and (19), note that for any operator $O$, 
\begin{eqnarray}
\int_{t^{\prime }}^{\infty } \frac{dt}{\tau} e ^{-(t-t ^{\prime})/ {\tau} } \int \frac{d^{n} x d^{n} p}{(2\pi \hbar ) ^{n}}  tr ( O  e ^{ \frac{i}{\hbar} (t-t ^{\prime} ) H } | \phi _{ {\bf x} , {\bf p} } \rangle \langle \phi _{ {\bf x}, {\bf p} } | e ^{ -\frac{i}{\hbar} (t-t ^{\prime} ) H } )
\end{eqnarray}
\[
=\int_{t^{\prime }}^{\infty } \frac{dt}{\tau} e ^{-(t-t ^{\prime})/ {\tau} } tr (O e ^{ \frac{i}{\hbar} (t-t ^{\prime} ) H } \underline{ \int \frac{d^{n}x d^{n}p}{(2\pi \hbar )^{n}} | \phi _{ {\bf x}, {\bf p}} \rangle \langle \phi _{ {\bf x}, {\bf p} }   | } e ^{ -\frac{i}{\hbar} (t-t ^{\prime} H ) } )
\]
\[
=\int_{t^{\prime }}^{\infty } \frac{dt}{\tau} e ^{-(t-t ^{\prime})/ {\tau} } tr (O e ^{ \frac{i}{\hbar} (t-t ^{\prime} ) H}  e ^{ -\frac{i}{\hbar} (t-t ^{\prime} H ) } ) \text{ (from Eq. (37))}
\]
\[
= tr O. \text{ (since $e ^{- \frac{i}{\hbar} (t-t ^{\prime} ) H }$ is unitary) }
\]
From the above equation we can prove Eq. (18) by taking $O$ as $| \phi _{ {\bf x} ^{\prime}, {\bf p} 
^{\prime} } \rangle \langle \phi _{ {\bf x} ^{\prime}, {\bf p} ^{\prime} } |$, and prove Eq. (19) by setting $O$ and $t ^{\prime}$ as $\rho (t_{0})$ and $t _{0}$, respectively. 

\section*{Appendix B}

Consider the case that the change of $V({\bf x})$ is insignificant in the
quantum scale and $m$ is very large. For the particle corresponding to the 
wave packet $| \phi _{ {\bf x} ^{\prime}, {\bf p} ^{\prime} } \rangle$ 
initially, it is known that the quantum effect can be ignored for
a long time. Hence the function $G({\bf x}, {\bf p}; {\bf x}^{\prime},
{\bf p} ^{\prime} ; \xi)$,  which is the probability to find this
particle around $({\bf x},{\bf p})$ at time $\xi$ under the Liouville flow, 
concentrates around the point $({\bf x},{\bf p}) =( {\bf X} ( {\bf x}
 ^{\prime}, {\bf p} ^{\prime}, \xi ), {\bf P} ({\bf x} ^{\prime}, {\bf p}
^{\prime}, \xi )) $. Here $({\bf X} ({\bf x}^{\prime}, {\bf p} ^{\prime}, \xi ),
{\bf P} ({\bf x} ^{\prime}, {\bf p} ^{\prime}, \xi))$ is the location of the
particle which originally is at $({\bf x} ^{\prime}, {\bf p} ^{\prime})$ after
passing time $\xi$ under the classical Liouville flow.

Approximating $G({\bf x}, {\bf p}; {\bf x}^{\prime}, {\bf p} ^{\prime} ; \xi) $ as 
$\delta ( {\bf x} - {\bf X} ( {\bf x} ^{\prime} , {\bf p} ^{\prime} , \xi) )
\delta ( {\bf p} - {\bf P} ( {\bf x} ^{\prime} , {\bf p} ^{\prime} , \xi) ) $
in Eq. (21), from Eq. (6) it is not hard
to see that $\langle \phi _{ {\bf x}, {\bf p} } | \rho (t) | \phi _{ {\bf x},
{\bf p} } \rangle (= r ( {\bf x}, {\bf p} , t) \tau )$ satisfies
\[
\frac{\partial}{\partial t}\langle \phi _{ {\bf x}, {\bf p} } | \rho (t) |
\phi _{ {\bf x}, {\bf p} } \rangle + \sum _{j=1} ^{n} \frac{p_{j}}{m} 
\frac{\partial}{\partial x _{j}} \langle \phi _{ {\bf x}, {\bf p} } | \rho (t) |
\phi _{ {\bf x}, {\bf p} } \rangle - \sum _{j=1} ^{n} \frac{\partial V ({\bf x}
)}{\partial x_{j}} \frac{\partial}{\partial p _{j} } \langle \phi _{ {\bf x},
{\bf p} } | \rho (t) | \phi _{ {\bf x}, {\bf p} } \rangle \simeq 0
\]
as $t>>t_{0}$. Taking $\langle \phi _{ {\bf x}, {\bf p} } | \rho (t) | \phi _{
{\bf x},{\bf p} } \rangle $ as the distribution function, therefore, the CTRW 
defined by Eqs. (5)-(8) can be related to the classical Liouville equation.

\section*{Appendix C}
Let
\begin{equation}
R({\bf x}, {\bf p},t)=\frac{1}{ \Omega _{1} \Omega _{2} T}
\int_{t}^{t+T}dt^{\prime }\int_{ | {\bf y} | < \Delta _{1}  }  d^{n}y 
\int _{|{\bf k} | < \Delta _{2} } d^{n}k \text{ }r({\bf x}+{\bf y},{\bf p}
+{\bf k},t^{\prime})
\end{equation}
be the macroscopic probability density of the CTRW. Here $T$ is a real
parameter so that for any two times $t_{1}$ and $t_{2}$ satisfying $
|t_{1}-t_{2}|<T$, they are indistinguishable macroscopically. Assume that
microscopically $r( {\bf x}, {\bf p} ,t)$ is homogeneous in the time scale 
of $T$ so that 
\begin{equation}
\frac{1}{T}\int_{t}^{t+T}dt^{\prime }\text{ }r( {\bf x}+{\bf y}, {\bf p}
+{\bf k},t ^{\prime}) \simeq R( {\bf x}, {\bf p},t)
\end{equation}
when $|{\bf y}|< \Delta _{1}$ and $|{\bf k}|< \Delta _{2}$.  
To show that $\Psi ( {\bf x}, {\bf p}; {\bf x}^{\prime}, {\bf p}
^{\prime };t-t^{\prime })$ in Eq. (23) can be taken as the macroscopic
transition rate of the CTRW, we need to prove that in a large scale
\begin{equation}
R({\bf x}, {\bf p},t) \simeq \int \frac{ d^{n}x^{\prime}
d^{n}p^{\prime }}{(2\pi \hbar )^{n}}\int_{t_{0}}^{t}dt^{\prime }\Psi (
{\bf x},{\bf p}; {\bf x}^{\prime },{\bf p}^{\prime };t-t^{\prime})
R( {\bf x}^{\prime }, {\bf p}^{\prime },t^{\prime })+S( {\bf x},
{\bf p},t)
\end{equation}
under Eq. (40), where $S( {\bf x}, {\bf p},t)$ is the macroscopic
source term. To prove that the above equation holds in a large scale,
we just need to insert Eq. (6) into Eq. (39) and note Eq. (40).

\end{document}